\documentclass[twocolumn]{aastex631}

\usepackage{amsmath}

\newcommand\hmi{{\it SDO}/HMI}

\begin{document}
\title{Observed Power and Frequency Variations of Solar Rossby Waves with Solar Cycles}
\author[0000-0003-2678-626X]{M. Waidele}
\author[0000-0002-6308-872X]{Junwei Zhao}
\affiliation{W. W. Hansen Experimental Physics Laboratory, Stanford University, Stanford, CA 94305-4085, USA}

\begin{abstract}
Several recent studies utilizing different helioseismic methods have confirmed the presence of large-scale vorticity waves known as solar Rossby waves within the Sun. 
Rossby  waves are distinct from acoustic waves, typically with longer periods  and lifetimes; and their general properties, even if only measured at  the surface, may be used to infer properties of the deeper convection zone, such as the turbulent viscosity and entropy gradients which are otherwise difficult to observe.
In this study, we utilize $12~$years of inverted subsurface velocity fields derived from the SDO/HMI's time--distance and ring-diagram pipelines to investigate the propoerty of the solar equatorial Rossby waves. 
By covering the maximum and the decline phases of Solar Cycle 24, these datasets enable a systematic analysis of any potential cycle dependence of these waves.
Our analysis provides evidence of a correlation between the average power of equatorial Rossby waves and the solar cycle, with stronger Rossby waves during the solar maximum and weaker waves during the minimum.
Our result also shows that the frequency of the Rossby waves is lower during the magnetic active years, implying a larger retrograde drift relative to the solar rotation.
Although the underlying mechanism that enhances the Rossby wave power and lowers its frequency during the cycle maximum is not immediately known, this observation has the potential to provide new insights into the interaction of large-scale flows with the solar cycle.
\end{abstract}

\section{Introduction} \label{sec:intro}
Rossby waves are commonly understood as planetary waves that arise in a planet's rotating atmosphere \citep{Rossby1939RelationBV}. 
On global scales, particle motions are subject to the pressure gradient force and the Coriolis force with the latter increasing with latitude. 
This creates spiral trajectories leading to large scale cells of vorticity that slowly propagate retrograde \citep{1985Hoskins, Dikpati_2020}.
Rossby waves are predicted by theory in rotating stellar atmospheres \citep{1978MNRAS.182..423P, 1981A&A....94..126P, 1982ApJ...256..717S, 1987AcA....37..341D} as well as in the solar atmosphere \citep{1986SoPh..105....1W}. 
Solar Rossby waves, also known as $r$ modes or inertial waves, cover a wide range of different types of waves that tend to have a characteristic angular frequency on the order of the solar rotation rate $\Omega$ \citep{Greenspan1972TheTO}.
Early observations of these types of waves were made using estimations from line-of-sight Doppler observations \citep{2001ApJ...560..466U} and periodic variations of the solar radius \citep{2015ApJ...804...47S}. 
Later, \citet{2017NatAs...1E..86M} reported a Rossby wavelike motion of magnetic bright points in the Sun’s corona. 

Direct observations, measuring the radial component of the vorticity $\zeta$, were only achieved recently \citep{2018NatAs...2..568L} using both the local correlation tracking method \citep{1988ApJ...333..427N} and the ring-diagram helioseismology method \citep{1988ApJ...333..996H}.
These observations described sectoral Rossby waves trapped in low-latitude regions.
Two-dimensional dispersion relation $\omega_{\ell m}$ of such waves can be derived by approximating the increase of the Coriolis force with latitude as linear and employing a shallow surface approximation \citep{1993JPO....23.1346V}, yielding
\begin{align}
    \omega_{\ell m} \approx -\frac{2m\Omega}{\ell(\ell+1)} \stackrel{\ell=m}{=} -\frac{2\Omega}{m+1},
    \label{eq:omega}
\end{align}
where harmonic degree $\ell$ equal to azimuthal degree $m$ limits the equation to only sectoral modes.
Later, more studies using different techniques confirmed this observation and revealed more details about the nature of the equatorial Rossby waves \citep{2019AaA...626A...3L, 2019ApJ...871L..32H, 2021arXiv210607251P}.
Other types of inertial waves have been discovered and discussed as well \citep{2021A&A...652L...6G, 2022NatAs...6..708H}.

Although \citet{2019AaA...626A...3L} demonstrated the capability of the time-distance helioseismology \citep{1993Natur.362..430D} to reveal equatorial Rossby waves, the study relied directly on travel-time measurements $\tau$. 
In this work, we further demonstrate how subsurface flow fields $\mathbf{v}$, the inverted data product of $\tau$ \citep{2012SoPh..275..375Z}, reveal Rossby wave signatures with higher spatial and temporal resolutions (compared to direct $\tau$ measurements). 

While by now, global properties of equatorial Rossby waves are well identified, their interactions with mechanisms that act on the Sun of similar scales, such as large-scale flows or magnetic fields, are not.
Rossby waves are thought to be excited within the convection zone, either within the tachocline or at supergranular depths \citep{2022ApJ...931..117D}, but their coupling with the solar dynamo, magnetic fields or flows are still unknown, although some coupling must happen directly or indirectly on some scales.
In fact, \citet{2019AaA...626A...3L} already hinted at a possible connection between Rossby wave amplitudes and the progression of the solar cycle. 
In this study, we will investigate this connection in greater detail.
The subsurface flow fields $\mathbf{v}$ are available for the period of 2010 \-- 2022, covering both the activity maximum and the declining phase of Cycle 24, allowing us to study the time dependence of the equatorial Rossby waves. 

This Letter is organized as follows: we describe our data-analysis procedure in Section~\ref{sec:data}, present our results on signatures of the equatorial Rossby waves and their dependence on the solar cycle in Section~\ref{sec:results}, and discuss these results and give our conclusions in Section~\ref{sec:conclusion}. 

\section{Data Analysis} \label{sec:data}

\subsection{Subsurface Flow Fields} \label{sec:flowfields}
The velocity fields $\mathbf{v}$ used in this study are obtained from the time-distance helioseismology pipeline \citep{2012SoPh..275..375Z} for Helioseismic and Magnetic Imager onboard Solar Dynamics Observatory \citep[\hmi;][]{2012SoPh..275..207S, 2012SoPh..275..229S}, with a duration of around $12~$years. 
The computation of $\mathbf{v}$ can be summarized as measuring $\tau$ through fitting cross-covariance functions that are calculated from tracked, remapped, and filtered Dopplergrams. 
Afterwards, travel-time differences are inverted for subsurface flow maps $\mathbf{v}$, using ray-path approximation kernels \citep{1996ApJ...461L..55K, 2001ApJ...557..384Z}.
The flow fields cover a horizontal range of $120^\circ\times120^\circ$ in $1026\times1026~$pixel$^2$ and span the depth from surface down to around $20$\,Mm, with a temporal cadence of $8~$hours.
We only use the shallowest layer, i.e., $0$ \-- $1$\,Mm, of the flow fields in this study.

Additionally, we prepare a set of ring-diagram helioseismology data, similarly obtained from the \hmi\ data analysis pipeline \citep{2011JPhCS.271a2008B} as a control sample. 
This dataset provides subsurface velocities during the same period as the time-distance data, but with a broader spatial sampling of $7.5^\circ~$pixel$^{-1}$ (derived from $15^\circ\times15^\circ$ tiles) and a larger temporal candence of $1/24$ of a synodic rotation ($\approx27.25$\,hr).

\subsection{Vorticity and Rossby waves} \label{sec:vorticity}
\begin{figure*}[htb]
    \plotone{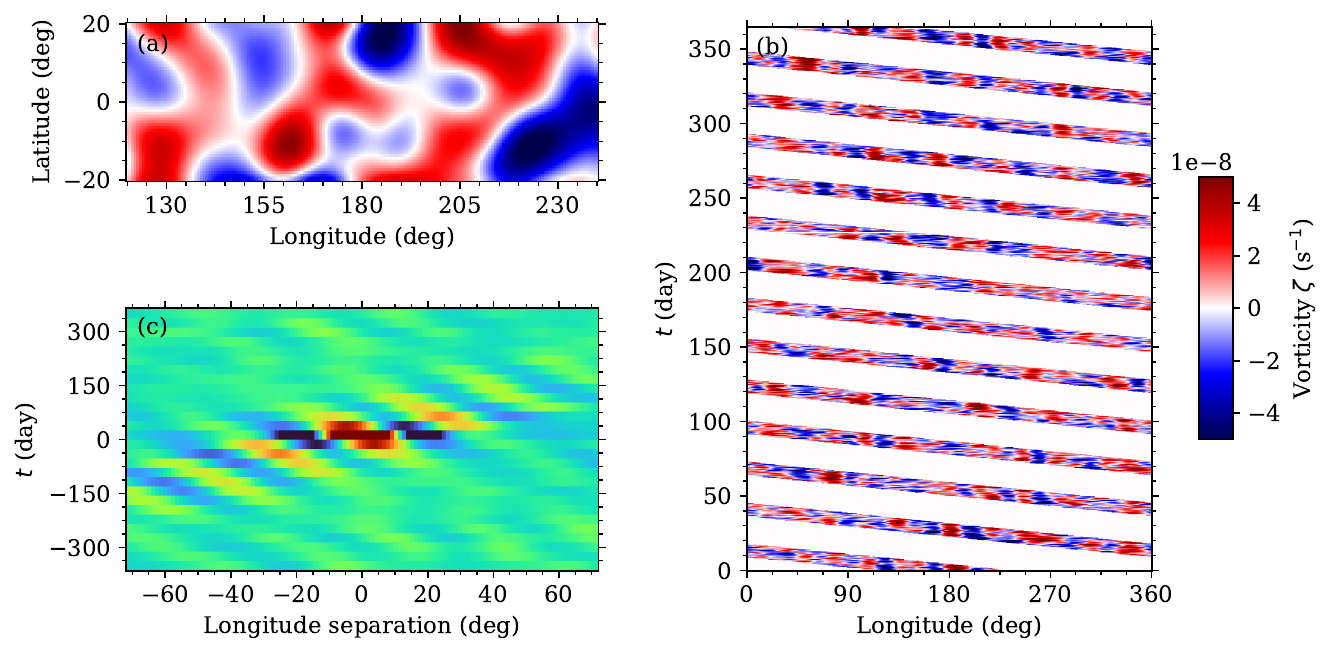}
    \caption{
        Vorticity patterns on the solar surface. (a): Vorticity $\zeta$ for an arbitrary point in time (after smoothing) as a function of longitude and latitude. (b): Vorticity tracked and stacked at $\theta=0\degr$ for $1~$year. Since subsurface flows are limited to $\phi\in[120^\circ, 240^\circ]$, the stack plot contains large horizontal gaps of $240^\circ$. The color scale is identical to that in panel a. (c): Cross-covariance computed for $\zeta$ averaged over a solar rotation as a function of time  and longitude separation. The color scale is arbitrary and dimensionless.
    }
    \label{fig:1}
\end{figure*}

Vortex-like patterns in the Sun can be identified by calculating the radial component of the vorticity $\zeta$, which is defined as \citep{201927...DCDS...39.6261}:
\begin{align}
    \zeta = \hat{\mathbf{r}}\cdot\mathbf{\nabla}\times \mathbf{v} = \frac{1}{r}\left( \frac{\partial v_\varphi}{\partial\theta} - \frac{1}{\sin\theta}\frac{\partial v_\theta}{\partial\varphi} + v_\varphi\cot\theta \right),
    \label{eq:zeta}
\end{align}
where $\hat{\mathbf{r}}$ is a unit vector pointing in the radial direction, $r$ is the radius, $\varphi$ the longitude, $\theta$ the latitude, and $v_\varphi, v_\theta$ the respective velocities. 
Once $\zeta$ is calculated, it is smoothed with a two-dimensional Gaussian function, with a standard deviation of $\sigma=7^\circ$, to remove small-scale features. 
Little Rossby wave power is expected beyond the Sun's mid-latitudes, we therefore limit the latitudinal extent of $\zeta$ to $\theta\in[-20^\circ, 20^\circ]$.
An example snapshot of $\zeta$, which is used in our follow-up analysis, is shown in Figure~\ref{fig:1}a. 

Since Equation~\ref{eq:omega} is given as a function of $\Omega$, we track $\zeta$ using the synodic rate $\Omega/2\pi=421.41~$nHz.
Stacking $\zeta$ with time $t$, as shown for $\theta=0\degr$ in Figure~\ref{fig:1}b, shows the tracking procedure and highlights slowly moving vorticity patterns, for which the most dominant longitudinal wavenumber is around $m=8$.
This vorticity pattern can be easily identified as a Rossby wave by computing cross-covariances as a function of both longitude separation and time delay (Figure~\ref{fig:1}c). 
This time-distance diagram of the Rossby wave shows clearly a retrograde phase speed and prograde group speed that is similar to the phase speed in magnitude.
When plugging $m=8$ into Equation~\ref{eq:omega}, a phase-speed of $v_{\rm p}=0.36^\circ~$day$^{-1}$ is obtained, and this is similar to the retrograde moving phase pattern seen in Figure~\ref{fig:1}c, propagating at roughly $v_{\rm p}=0.33^\circ~$day$^{-1}$.

\subsection{Decomposition} \label{sec:decomp}
Identifying equatorial Rossby waves can be done by calculating cross-covariances as demonstrated in Figure \ref{fig:1}c, it is however convenient to transform the data into  Fourier domain and express wave properties using the azimuthal degree $m$ and the temporal frequency $\nu = \omega / 2\pi$. 
Thereby, $\zeta(\theta, \varphi)$ is decomposed into spherical harmonics $Y_{\ell, m}$ yielding
\begin{align}
    \widetilde{\zeta}_{\ell, m} = \sum_{\theta, \varphi}\zeta(\theta, \varphi)Y^*_{\ell, m}(\theta, \varphi)\sin\theta\,.
    \label{eq:decomp}
\end{align}
To avoid side-lobe patterns as much as possible, we apply spatial Hann taper-windows \citep{1958DoverPubl} to $\zeta(\theta, \varphi)$. 
Afterwards, the power is calculated as 
\begin{align}
    p_{\ell, m}(\nu) = \left|\sum_t\widetilde{\zeta}_{\ell, m}(t)\exp{(\mathrm{i\,} 2\pi\nu t)}\right|^2\,,
    \label{eq:power}
\end{align}
whereas we limit ourselves to sectoral modes with $\ell=m$ in the following, due to non-sectoral modes being surpressed \citep{1998ApJ...502..961W, 2000ApJ...529..997Y} in the Sun.
Only a fraction of the solar surface is covered in $\zeta$, leading to large gaps in the tracked $\zeta(\theta, \varphi, t)$. 
When calculating $p_{\ell=m, m}(\nu) = p_{m}(\nu)$, these gaps will manifest as power leaking from the sectoral modes $(m, \nu)$ into $k$-order side-lobes $(m\pm k, \nu \mp k\Omega/2\pi)$ with integer $k$ \citep[see][for an in-depth explanation]{2019AaA...626A...3L}.
As a consequence, the power spectrum is limited to approximately $\nu\in[-300, 50]~$nHz, even though a cadence of $8\,$hours allows a much larger frequency range. 
Because of this, after tracking $\zeta$, we average 15 timesteps corresponding to $5~$days, reducing the cadence from $8~$hours to $5~$days, which also significantly decreases the noise level in $\zeta$ and thus in $p_m(\nu)$.

\section{Results} \label{sec:results}

\subsection{Detection of Rossby Waves} \label{sec:wavepower}
\begin{figure*}[htb!]
    \plotone{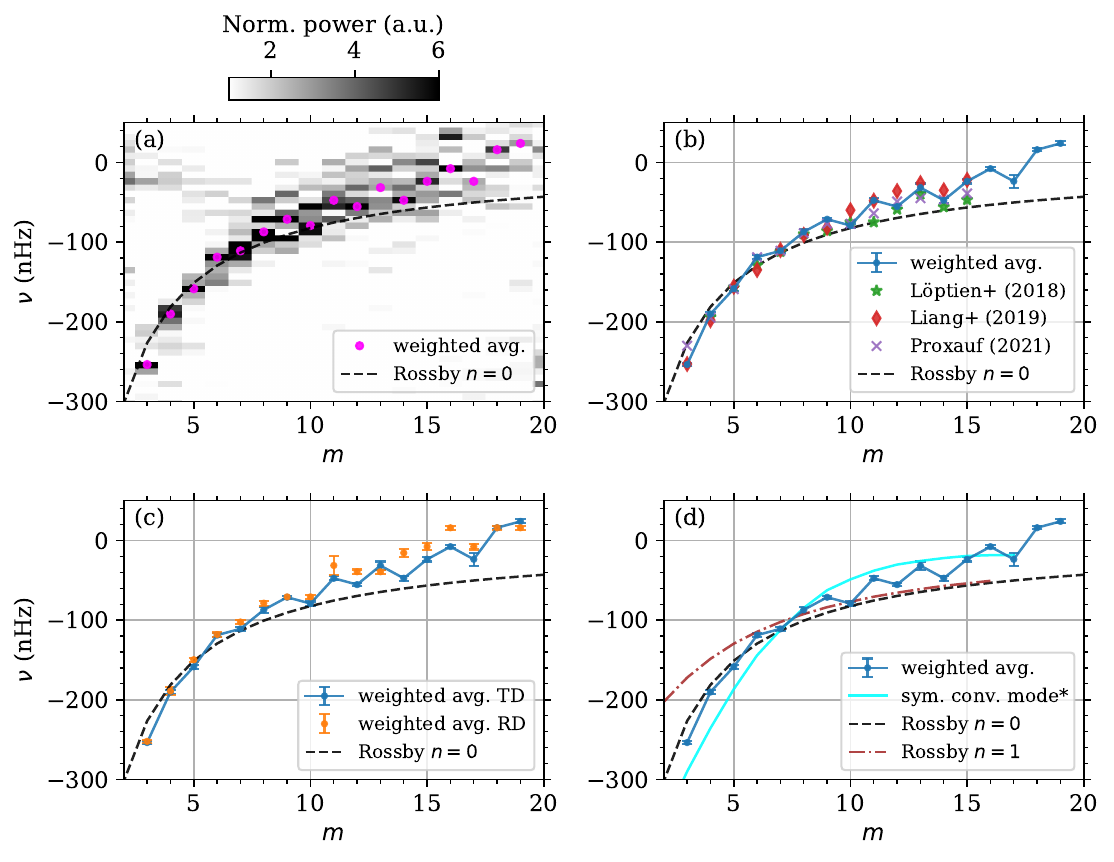}
    \caption{
        Estimates of $p_m(\omega)$ and $\nu_m$ using time-distance (TD) and ring diagram (RD) helioseismic data. 
        (a): normalized power $p_m(\omega)$ with $\nu_m$ overplotted for TD data (magenta  dots). Theoretical dispersion relation (Eq. \ref{eq:omega}) is shown as a black dashed curve. (b): Comparison of $\nu_m$ from this study with those from selected studies. Errorbars for those studies are not shown for display purposes. (c): Results for $\nu_m$ using TD data (blue dots connected by line) versus using RD data (orange dots). (d): Theoretical curves of different possible power sources in the $\nu\in[-300, 50]~$nHz interval \citep[dispersion relations are taken from][]{2022A&A...662A..16B}. The brown, dash dotted curve denotes the $n=1$ Rossby wave dispersion relation, while the cyan line shows the ($k=1$-order sidelobe) symmetric convective mode.
        \label{fig:2}
    }
\end{figure*}

Power contained in a sectoral Rossby wave with the azimuthal oder $m$ is given as the amplitude of $p_m(\nu)$. 
This is shown in Figure \ref{fig:2}a, where a ridge can be seen that follows approxiamtely the dispersion relation given in Equation \ref{eq:omega}. 
The exact frequency locations \mbox{$\nu_{\ell=m, m} = \nu_m$} are then determined for each $m$ using a weighted average:
\begin{align}
    \Bar{\nu}(\nu_0) &= \frac{1}{N_\nu}\sum_\nu p_m(\nu) L(\nu_0, \gamma, \nu)\,,\\
    \nu_m &= {\rm max}\left[\Bar{\nu}(\nu_0)\right]\,,
\end{align}
where $N_\nu$ is the amount of data points and $L(\nu_0, \gamma, \nu)$ is Lorentzian kernels centered at $\nu=\nu_0$ with the width $\gamma=25\,$nHz.
Errorbars $\sigma_\nu$ are derived from the statistical variance
\begin{align}
    \sigma_\nu^2 &= \frac{1}{N_\nu}\sum_\nu(\nu - \mu)^2\,,\\
    \mu &= \frac{\sum_\nu\left[\nu_m L(\nu_0=\nu_m)\right]}{\sum_\nu L(\nu_0=\nu_m)}\,.
\end{align}
As mentioned, the $m=8$ mode is dominant on the Sun. 
For display purpose, for each $m$ the power spectrum is divided by the average power in the aforementioned $\nu\in[-300, 50]~$nHz interval.

To gauge the performance of the time-distance flow fields, a comparison with selected previous studies is shown in Figure \ref{fig:1}b.
Qualitatively, the results agree with each other, although the resulting dispersion relation derived here appears noisier.
The similar behavior can be seen in the dispersion relation we calculated from the ring-diagram velocities (Figure \ref{fig:2}c). 
It is likely that a difference in the determination of $\nu_m$ causes such a discrepancy  \citep[Note that weighted averages are used here while Lorentzian fitting was used in the previous studies by][]{2018NatAs...2..568L, 2019AaA...626A...3L, 2021arXiv210607251P}.
Nevertheless, all studies show a shift away from the theoretical dispersion relation toward larger $\nu$. 
While approximations used to derive Equation \ref{eq:omega} may result in an underestimate of the true dispersion relation, physical causes may be at play. 
Possible candidates, as modeled by \citet{2022A&A...662A..16B, 2022A&A...666A.135B} and shown in Figure \ref{fig:2}d, include the higher radial order ($n=1$) Rossby wave, as well as a hypothetical dispersion relation of solar columnar thermal modes, labeled in the figure as symmetric convective modes with $\ell=m$.
Note that the latter is expected to appear in the prograde $\nu$-range, although the $k=1$-order sidelobe would exhibit similar frequencies to those given in Equation~\ref{eq:omega}. 

Although not shown here, the $\lvert m \rvert = \ell - 1$ spectrum, a weak and noisy but nevertheless significant power ridge is seen at frequencies similar to those reported in \citet{2022NatAs...6..708H}.
No considerable power is found for $\lvert m \rvert < \ell - 1$.

\subsection{Variations of Rossby Waves with the Solar Cycle} \label{sec:solarcycle}
Despite the increase of uncertainties in the frequency determination, two years of data are sufficient to adequately resolve sectoral Rossby waves in $p_m(\nu)$ spectra. 
Thus, the $12~$years of data used here are divided into segments of two years. 
This would result in a total of six data points, which is then increased to 11 by allowing a one year overlap between the segments.

\subsubsection{Wave Power} \label{sec:solarcycle_wavepower}
\begin{figure*}
    \plotone{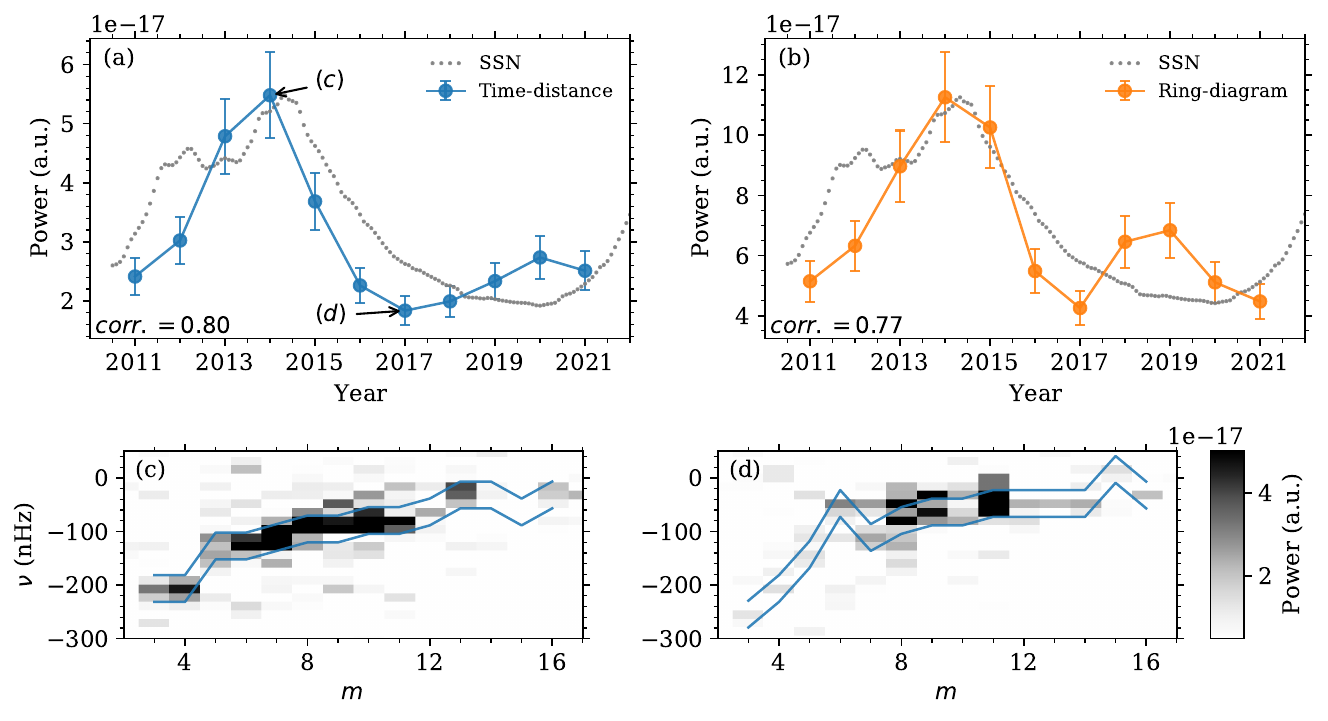}
    \caption{
        Average Rossby wave power as a function of time. Each data point in both top panels represents the average power extracted from a two-year segment with one year overlapping between neighbouring data points. (a): Comparison of average wave power, derived from the TD data, with the monthly smoothed sunspot number (SSN, normalized to fit the scale). Annotations \textit{(c)} and \textit{(d)} indicate the respective powers that are extracted from the power spectra shown in panels c and d. The value $corr.$ denotes the sample correlation coefficient between the wave power and the SSN. (b): Same as panel a but for the RD data. Note the difference in the vertical scale. (c): Power spectrum $p_{m}(\nu)$ during the cycle maximum in 2014. Blue lines located at $\nu_{m}\pm\nu_{\rm avg}$ indicate the frequencies between which power is averaged to calculate a data point in panel a. (d): Same as panel c but for the power minimum (not cycle minimum) in 2017.
        \label{fig:3}
    }
\end{figure*}

After performing the same procedure demonstrated in Sections~\ref{sec:vorticity} \& \ref{sec:decomp} to obtain $\nu_m$ for each segment, the power in $\nu_m\pm\nu_{\rm avg}$ is averaged with $\nu_{\rm avg}=20\,$nHz for all the $m\in[3, 16]$.
This results in a time dependent power curve, shown in Figure~\ref{fig:3}a.

Errorbars are estimated using a mathematical procedure: Assuming that the power $p_m(\nu)$ is estimated by the periodogram as in Eq.~\ref{eq:power}, its variance adheres to a $\chi_2^2$ distribution with ${\rm var}=4p_m(\nu)^2$. 
Apodizing the time-series $\widehat{\zeta}(t)$ and averaging $p_m(\nu)$ over $N_m$ and $N_\nu$ data points reduce the error $\sigma_p$ to \citep{1981book...priestley:errors}:
\begin{align}
    \sigma_p^2 = \frac{4\lambda^2}{N_mN_\nu}p^2\,,
    \label{eq:power_error}
\end{align}
where $\lambda\approx0.86$ accounts for the effect of apodizing the data (using a Hann window) and $p$ is the average power.

Overplotting the monthly smoothed sunspot number \citep[SSN, data from][]{SWPC_SSN} on the Rossby wave power reveals a high correlation between the two, with the Rossby wave power showing its peak near 2014 during the solar activity maximum. 
Time-distance helioseismic data products are known to be strongly affected by surface magnetic fields \citep{2023ApJ...950...63M}, such that it can be argued that the subsurface flow fields used in this work are affected by surface magnetic field, causing this cycle-dependent power variations in Rossby waves. 
Using ring-diagram helioseismic data, which are presumably less affected by the surface magnetic field, a similar correlation with the solar cycle can still be seen (Figure~\ref{fig:3}b). 
Although systematic effects due to surface magnetism can not be easily outruled, it is likely that Rossby wave power does show a solar cycle dependence. 
The wave power computed from the RD data is roughly $2~$times stronger than the TD data result. 
This discrepancy stems from the treatment of the velocity data, since RD data is limited to $15^\circ$ tiles, while the TD velocities are smoothed using a Gaussian, resulting in slightly different values.
Even small differences between velocity amplitudes propagate through calculations and get amplified in the final estimation of power (in Eq.~\ref{eq:power}). Panels c and d of Figure~\ref{fig:3} show the two-year segment power spectra $p_m(\nu)$ during the solar maximum (2013 \-- 2015) and minimum years (2016 \-- 2018), respectively. 

\subsubsection{Frequency Location} \label{sec:solarcycle_freqloc}
\begin{figure*}
    \plotone{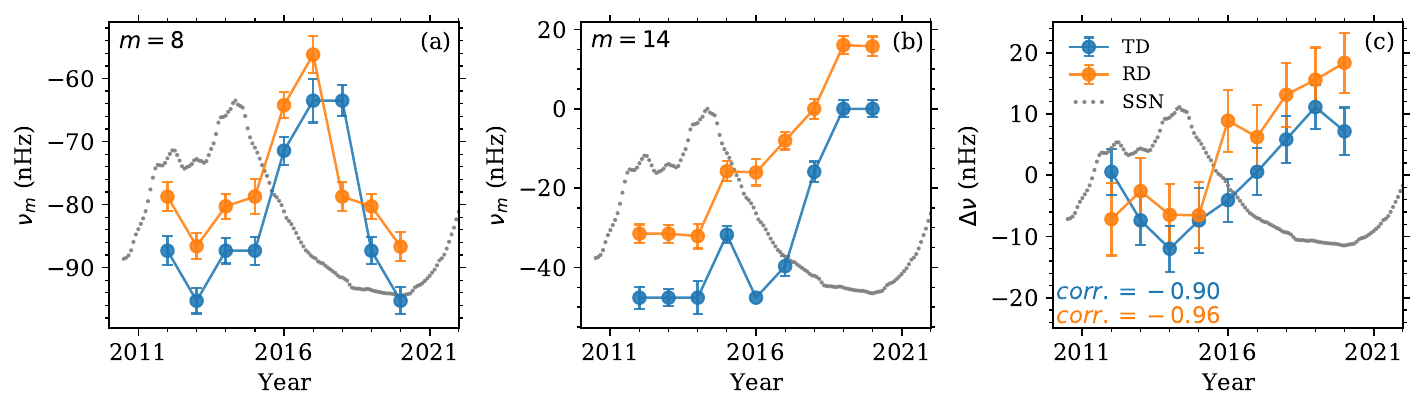}
    \caption{
        Frequency locations $\nu_m$ (for selected $m$) and average frequency deviation $\Delta\nu$ shown as functions of time. (a): Frequency location $\nu_{m=8}$ for TD data (blue) and RD data (orange). Similar to Figure~\ref{fig:3}, the monthly SSN is shown as grey dots. (b): Same as panel a but for $\nu_{m=14}$. (c): Average frequency deviation $\Delta\nu$ obtained from the TD and RD data.
        \label{fig:4}
    }
\end{figure*}

Aside from the variation in power of the Rossby waves, it appears that both the noise level and the frequency locations $\nu_m$ of the waves vary with time. 
Qualitatively, this can be seen in Figure~\ref{fig:3}c and \ref{fig:3}d.
In the case of $\nu_m$, performing an average is not straightforward, so we tackle this in two ways. 
First, we increase the total duration of each segment to four years, thereby increasing the amount of overlap between each segment to two years (reducing the number of data points to 9).
Second, an average frequency deviation $\Delta\nu$ can be defined as the variation of $\nu_m$ from a reference $\nu_m^{\rm ref}$ with 
\begin{align}
    \Delta\nu = \frac{1}{N_m} \sum_m \left( \nu_m - \nu_m^{\rm ref} \right) \,,
    \label{eq:deltanu}
\end{align}
where we use the $\nu_m$ derived from the full $12~$year power spectrum (as shown in Figure~\ref{fig:2}) as the reference $\nu_m^{\rm ref}$. 
The time-dependent curves of $\nu_m$ and $\Delta\nu$ are plotted in Figure~\ref{fig:4}.

For both Figure~\ref{fig:4}a and \ref{fig:4}b, a general trend towards larger $\nu$ as the cycle declines can be seen, although larger $m$ appear to retain large $\nu$ while medium $m$ recede to lower $\nu$ again. 
This general trend can be seen in Figure~\ref{fig:4}c where the average $\Delta\nu$ shows an increase as the cycle declines for both results from the TD and RD data.
Overall, the behavior of $\Delta\nu$ demonstrates that Rossby waves tend to propagate retrograde with a faster speed during the solar activity maximum.

\subsubsection{Latitudinal Dependency} \label{sec:solarcycle_latdep}
\begin{figure*}
    \plotone{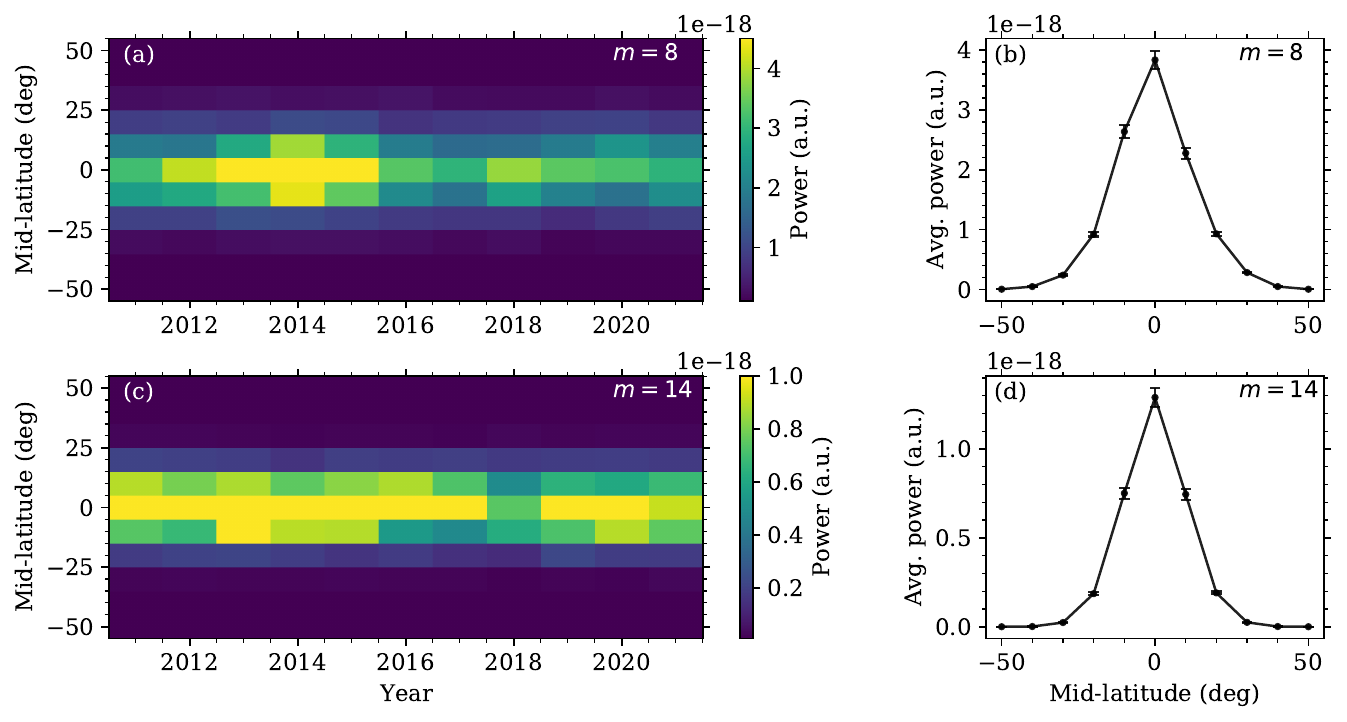}
    \caption{
        Rossby power as a function of time and mid-latitude. Each latitudinal segment is $20^\circ$ in size and has a $10^\circ$ overlap with neighbouring segments. (a): Power average for $m=8$. (b): Further average over time, indicating the mean power distribution per latitudinal segment. (c): Same as panel a but for $m=14$. (d): Same as panel b but for $m=14$.
        \label{fig:5}
    }
\end{figure*}

With Rossby wave power showing amplification (or suppression) and the phase speed exhibiting (de-) acceleration during the progression of the solar cycle, it remains unclear whether such effects are caused by the configuration of the global magnetic field, or by modifications of local effects due to solar activity. 
A possible analysis to reveal any local effects is to further divide the data into latitudinal segments to spatially resolve the waves' power distribution. 
We define segments of $20^\circ$ in size ranging from $\theta\in[-60^\circ, 60^\circ]$ with $10^\circ$ of overlapping between neighboring segments (resulting in 11 data points in latitudinal direction).
At high latitudes, the differential rotation of the Sun becomes significant and the tracking rate $\Omega$ must be modified \citep{2022NatAs...6..708H} using:
\begin{align}
    \Omega^\prime = \Omega - \left[ 54.39\sin^2(\theta) + 75.44\sin^4(\theta)\right]2\pi{\rm~nHz}\,,
    \label{eq:difftrack}
\end{align}
where we use the mid-latitude for $\theta$ for each segment.
It must be noted that, by using a modified tracking rate $\Omega^\prime$, the dispersion relation Equation~\ref{eq:omega} also changes. 
A simple approach to account for this is to average the entire spectrum $p_m(\nu)$ over $\nu$ for a single $m$, as the resulting average power does not depend on any frequency shifts introduced by a different tracking rate $\Omega^\prime$.
The resulting latitude-time diagrams are shown in Figure~\ref{fig:5} for $m=8$ and $m=14$.

Since Rossby waves are expected to be confined near the equator, almost no power is found at  latitudes $\theta>20^\circ$. 
While the trend of solar-cycle dependence can be seen in these plots, no strong latitudinal dependence is found. 
For $m=8$, the southern hemisphere contains slightly more power, which may coincide with more magnetic activity occurring in the south during Cycle 24, but this is not significant. 

\section{Discussion \& Conclusion} \label{sec:conclusion}
Following a data-analysis process similar to previous authors \citep{2018NatAs...2..568L, 2019AaA...626A...3L, 2021arXiv210607251P, 2021A&A...652L...6G, 2022NatAs...6..708H} but using different types of data, we demonstrate the capability of time-distance subsurface flow fields of investigating the nature of the solar equatorial Rossby waves. 
The long-term behavior of Rossby wave power is well correlated with solar activity (Figure~\ref{fig:3}), and both time-distance and ring-diagram data show that the Rossby wave power is stronger during the Sun's magetic active phase and weaker during the quiet period. 

What causes the Rossby waves to be stronger in the magnetic active phase? 
Excitation of the equatorial Rossby waves is not conclusively understood, although it is suspected that inverse cascading from supergranular energy is the main driving mechanism \citep{2022ApJ...931..117D}. 
Supergranular energy being modified during the solar cycle could therefore serve as a mechanism of equatorial Rossby wave enhancement. 
It would be of interest to investigate such a dynamic in a suitable 3D simulation using, e.g., Rayleigh code \citep{2016ApJ...818...32F}.

Although we cannot fully exclude the possibility that the increase of Rossby wave power with magnetic activity is caused by systematic effects in the velocity measurements, a decrease in $\Delta\nu$ (Figure~\ref{fig:4}c) supports the otherwise. 
Active regions on the solar surface are known to move slightly faster due to their anchorage in deeper faster rotating layers \citep{1983ApJ...270..288S}. 
If the Rossby wave power presented here is coupled to systematic effects stemming from surface magnetism, it would also be expected an increased prograde speed for the same Rossby waves during increased solar activity. 
However, as presented, we find the opposite.

The equatorial Rossby waves show a trend of {\it increased retrograde} speed during the cycle maximum, although the trend is not consistent across all $m$: Frequency deviations found in $\Delta\nu$ (Figure~\ref{fig:4}c) exhibit a strong anti-correlation, which weakens considerably for $m=8$, potentially due to a phase-shift between $\nu_8$ (Figure~\ref{fig:4}a) and the cycle activity. 
Such a behavior can be explained by the interaction of Rossby waves with magnetic fields. 
Magnetic Rossby waves split onto a fast and a slow branch, with the former showing more retrograde frequencies as the strength of the solar toroidal field grows \citep{Dikpati_2020}. 
Slow branch magnetic Rossby waves are expected to exhibit slow, prograde frequencies during cycle maximum, which we do not find in $p_m(\nu)$. 
Another conceivable origin could be torsional oscillations. 
While a cycle dependency of the torsional oscillations shows frequency variations of less than $1~$nHz \citep{2022A&A...664A...6F}, it must be considered that the latitudinal range selection of $\theta\in[-20^\circ, +20^\circ]$ may introduce a time dependency through truncating the solar activity belt:
During the later stages of the cycle, active regions are more likely to be found within the latitudes of $\theta\in[-20^\circ, +20^\circ]$, resulting in a plasma acceleration toward the end of the cycle \citep{2009LRSP....6....1H}. 
Translating $\Delta\nu$ found in Figure~\ref{fig:4} into phase-speed, we find an average deviation of $\Delta v_{\rm p} \approx 11~$m\,s$^{-1}$ (for $m=8$). 
The zonal flow component introduced by a migrating activity belt has an amplitude of roughly $5~$m\,s$^{-1}$, which is slower than $\Delta v_{\rm p}$ but of the similar order of magnitude.

The latitudinal dependency (Figure~\ref{fig:5}) has the potential to show power enhancement for either the northern or southern hemisphere, although the resolution of the average power is poor. 
In fact, the southern hemisphere shows slightly more power around the solar maximum in 2014. 
For $m=8$ the local power increase is significant enough to cause a weak power inequality between both hemispheres that persists throughout the cycle progression (Figure~\ref{fig:5}b).
Such an inequality may indicate that the coupling of the Rossby wave power happens locally, e.g., in this case to the southern portion of the interior toroidal field.

In conclusion, we demonstrate a Rossby wave power variation that follows the progression of the solar cycle, using time--distance and ring-diagram subsurface flow data. 
The average frequency deviation $\Delta\nu$ of the equatorial Rossby modes show a tendency toward faster retrograde motion during the solar cycle maximum, demonstrating the coupling between equatorial Rossby waves and solar cycles. 
In turn, theoretical studies regarding the underlying mechanism of this coupling process can enable new techniques using Rossby wave parameter measurements to learn about, e.g., the deep magnetic field. 

SDO is a NASA mission, and HMI is an instrument developed by Stanford University under a NASA contract. We thank M. Dikpati and S. P. Rajaguru for reviewing the manuscript as well as other colleagues from the Stanford solar group for helpful discussion.

\bibliography{lit}{}
\bibliographystyle{aasjournal}

\end{document}